\documentstyle[12pt,epsf]{article}
\renewcommand{\bar}[1]{\overline{#1}}

\renewcommand{\mathrm}{{\rm}}

\def\pslash{\hbox{p\kern-.5em\lower.3ex\hbox{/}}}
\def\ru1{\rule[-0.4truecm]{0mm}{1truecm}}

\begin{document}


\bigskip\bigskip
{\centerline{\Large \bf Peripheral Meson Model of Deep 
Inelastic}} 
{\centerline{\Large \bf 
Rapidity Gap Events
}}

\vspace{30pt}

\centerline{
\bf Hung Jung Lu$^{a}$, Rodrigo Rivera$^{b}$ and Ivan 
Schmidt$^{b}$$^{\dagger}$}

\vspace{8pt}
{\centerline{$^a$Knowledge Adventure Inc., 1311 Grand Central 
Ave., Glendale, CA 91201, U.S.A.}}

\vspace{8pt}
{\centerline {$^b$Departamento de F\'\i sica,
Universidad T\'ecnica Federico Santa Mar\'\i a,}}

{\centerline {Casilla 110-V, Valpara\'\i so, Chile}}

\vspace{8pt}

\centerline{$^\dagger$e-mail: ischmidt@fis.utfsm.cl }


\vspace{60pt}
\begin{center} 
{\large \bf Abstract}
\end{center}

We show that a peripheral meson model can explain the large 
deep inelastic electron-proton scattering rapidity gap events 
observed at HERA.

\vfill
\centerline{
PACS numbers: 13.60.Hb, 14.40.-n}
\vfill

\newpage

\section{\bf Introduction}

Consider a proton at rest. Surrounding this proton there is a 
cloud of mesons ($\pi, \rho, \omega, \phi$, f2, etc...), which 
is fairly diluted at a distance that is large compared to 
the proton radius. That is, every meson is well separated from 
the rest of the mesons. Furthermore, due to the low density of 
mesons, nuclear models of proton-meson interactions should work 
on this regime, making perturbation theory a valid approximation,
because the expansions are not only on the powers of the 
proton-meson effective coupling, but the series is also 
suppressed by powers of the product between the meson mass and 
the physical meson-proton distance\cite{Meson}.

Now let us imagine a high $Q^2$ virtual photon in deep inelastic 
scattering, which due to the high $Q^2$ has vanishing size. 
Sometimes this photon will collide with the proton core, which 
constitutes a typical deep inelastic scattering events.
but in other instances it scatters off one of the isolated 
mesons in the disperse meson cloud.
The photon then breaks down the meson, and the pieces of the 
broken meson fragment independently of what happens to the 
proton. In fact, if the meson is far away, and if for instance 
the meson is a neutral pion, the most probable 
outcome is that the proton core will not be affected by the 
interaction between the photon and the meson, and after 
the meson is broken, the proton core will maintain its identity 
as a proton. Of course, if on the other hand the meson is 
charged, or if the core suffers the effects from the hard 
interaction, the proton can get excited into final states such 
as neutron, a $\Delta$(-,0,+,++), etc.

The basic picture is then the following: after the interaction we
have a broken meson and a baryon which basically does not move 
much. Now, let us boost this picture to the laboratory system of 
HERA. What we see is that the proton looses very little momentum 
and continues down the beampipe, and the meson fragments are 
observed in the central rapidity zone, with a rapidity gap 
between the meson fragments and the proton direction (forward 
direction of beampipe).

This picture we just presented gives a natural simple explanation 
of the rapidity gap events observed at HERA\cite{Zeus95}. It is 
the purpose of this paper to show that this is indeed the case, 
doing the corresponding calculation in detail.

The peripheral meson model we are considering here is different 
from the Pomeron exchange model\cite{W97}, which is the one 
that is usually ascribed to these large rapidity gap events. In 
fact, these two models give quite different predictions. 
Specifically the meson model predicts the presence of several 
excited baryonic states, besides the proton, in the forward 
direction of HERA; while the Pomeron model predicts that only the
proton will be present there. Therefore future measurements in 
these region should be able to clearly distinguish between the 
two models.

\section{\bf The Model}

The model we are considering is based on the diagram shown in 
Fig. 1, where a rapidity gap appears between the baryonic 
beampipe system $N^*$ and the hadrons comprising system $X$.

\vspace{0.5cm}
\begin{figure}[htb]
\begin{center}
\leavevmode {\epsfysize=10cm \epsffile{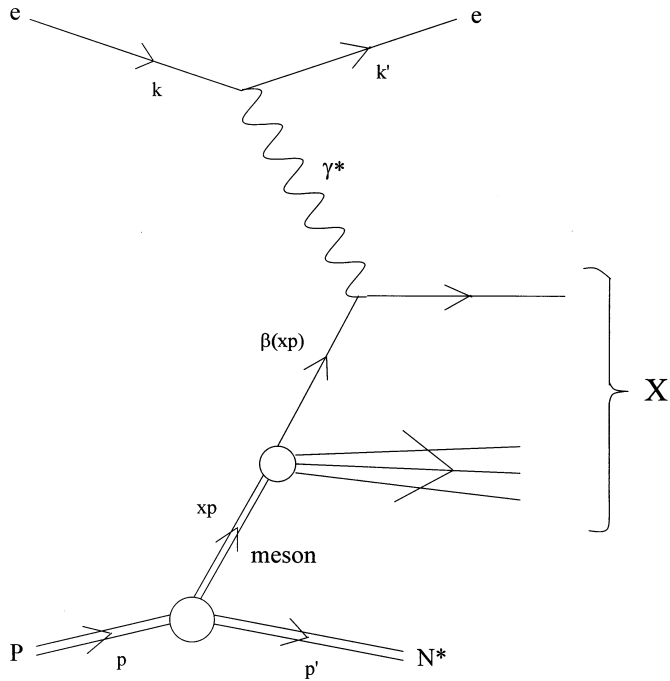}} 
\end{center}
\caption[*]{\baselineskip 13pt 
Diagram for the process $e(k)+p(p) \to e(k')+N^*(p')+X$, in which
a meson produces a rapidity gap between $N^*$ and $X$.} 
\label{usmth76f1} \end{figure}
 
The measured diffractive structure function is in general a 
function of three variables, which can be taken as the photon 
momentum transfer $Q^2$, the Bjorken variable $Q^2/2p\cdot q$, 
and the variable $\beta=Q^2/2q\cdot(p-p')$. Since the cross 
section is dominated by small $t=(p-p')^2$ values ($\arrowvert 
t\arrowvert \ll Q^{2},M_{X}^{2}$), both $\beta$ and the 
variable $x =q\cdot (p-p')/p\cdot q$ can be reconstructed  from
the measured quantities $x_{Bj}=x\beta$ and 
$M_{X}^{2}/Q^{2}=(1-\beta )/\beta $.

Therefore the whole process depicted in Fig. 1 implies that the 
diffractive structure function can be expressed as:
\begin{equation}
F_{2}^{d}(x_{Bj},Q^{2},\beta )=\sum _{m}f_{m/p}(x ,Q^{2})F_{2}^{m}(\beta
,Q^{2}).
\end{equation}
Here $f_{m/p}(x ,Q^{2})$ is the probability of finding the 
emitted meson carrying a fraction $x$ of the proton momentum, 
and $F_{2}^{m}(\beta,Q^{2})$ is the usual deep inelastic 
structure function of the corresponding meson. Then the 
photon scatters on a quark that carries a fraction 
$\beta$ of the meson momentum. Although in principle the sum is 
over all mesons, we will see below that a limited number actually
contributes to the process. Notice that we are in fact using an 
equivalent particle approximation, in which the process 
is dominated by the region in which the particle (in this case 
the meson) is close to its mass shell.

Since we are not looking into spin effects, for simplicity we 
will assume that the partonic distributions inside a meson are 
not very different form those inside a pion\cite{GRV92}.
Therefore our task is to work out expressions for the 
different distribution functions of mesons inside a 
proton.

The equivalent meson approximation allows for a separation of 
the amplitude for the whole process into an amplitude for a 
transition to a baryon $N^*$ and a near mass-shell meson, 
followed by the amplitude for the interaction of the meson with a
particle $a$ (in our case the off-shell photon). Using this it is
easy to show that the external and internal cross sections are 
related through:
\begin{equation}
d\sigma _{ap}=f_{m/p}(x,\tilde K^2)\ dx\ d\sigma_{a\pi },
\end{equation}
where the meson distribution inside a proton is then:
\begin{equation} 
f_{m/p}(x,\tilde K^{2})=\intop _{x^{2}M_{p}^{2}\over 
1-x}^{\tilde K^{2}}{x\over 16\pi ^{2}}{\mid M_{mp}\mid ^{2}\over
(K^{2}+m_{m}^{2})^{2}}dK^{2}.
\end{equation}  
Here $\mid M_{mp}\mid ^{2}$ is the squared unpolarized amplitude 
for meson-$N^*$ emission (summed over final spins of the meson, 
and averaged over initial spins of the proton),  $t=-K^2$ is 
the momentum squared of the pion, and $m_m$ and $M_p$ are the 
meson and proton masses respectively. $\tilde K^2$ is a 
momentum cut-off scale inherent to the formalism of 
equivalent particle approximation, and in our case it 
represents a parameter that fixes the separation of the pion 
from the color field of the proton. So we expect that its value 
be of the order of $\Lambda _{QCD}$ or less, although in our 
calculation we will leave it as an adjustable parameter. Notice 
that, as expected, the integral is dominated by small $t$ values.

The amplitude $M_{mp}$ contains a baryon-meson-baryon form 
factor, which can be taken as\cite{BH89}:
\begin{equation} 
F(K^{2})=\left(\Lambda ^{2}-m_{m}^{2}\over \Lambda 
^{2}+K^{2}\right)^{n},
\end{equation} 
where $\Lambda$ is a parameter and $m_{m}$ is the meson mass. 
In the parameterization of Ref. [5], the exponent $n$ is equal 
to $1$, except for the $\rho N\Delta$ case, where 
$n=2$.\cite{BH89}

\subsection{\bf Pion distribution inside the proton:}
We start with the lightest meson, the pion, in which case our 
general formula is exact. The amplitude for 
the whole process $M_{ap}$ is related to the pion-$a$ subprocess 
through:
\begin{equation} 
iM_{ap}=<p'\mid -ig_{\pi NN}\gamma _{5}\mid p>{i\over t-m_{\pi
}^{2}}iM_{a\pi }F_{\pi NN}(K^2),
\end{equation} 
where $p$ and $p'$ are the initial and final proton momenta, 
$g_{\pi NN}$ is the proton-pion coupling constant, and $F_{\pi NN}(K^2)$ is the 
proton-pion form factor. We are here considering the $\pi_0$ 
case, and for $\pi_{\pm}$ we simply need to multiply $M_{ap}$ by 
an isospin factor of $\sqrt{2}$. Then after squaring this result 
we get: \begin{equation} 
\mid M_{\pi pp}\mid ^{2}=g_{\pi NN}^{2}Q^{2}\mid F_{\pi 
NN}\mid ^{2}.
\end{equation}
Hence the pion distribution inside a proton is then:
\begin{equation} 
f_{p\to \pi_{0}p}(x,\tilde K^2)=x\intop _{x^{2}M_{p}^{2}\over 
1-x}^{\tilde K^2}{\alpha _{\pi NN}\over 4\pi }{K^2\over 
({K}^{2}+m_{\pi }^{2})^{2}}\big\arrowvert F_{\pi 
NN}(K^2)\big\arrowvert ^{2}dK^2,
\end{equation}
where $\alpha _{\pi NN}=g_{\pi NN}^{2}/4\pi$. 
At small $x$, the pion distribution is then 
proportional to $x$. That is, it will have a
limited contribution for the events of our interest. 
The same is true for the transition $p\to \pi\Delta$, and 
also for and other scalar or pseudoscalar meson.

\subsection{Vector meson distributions:}
Here we will study the vector meson distributions, and show that 
these are the most important contributions at the small $x$ 
region of our interest. In fact, most of the effect comes from 
the omega meson (proton in the final state), and the rho 
(nucleon and delta isobars in final state).

We will need to generalize the equivalent photon 
approximation\cite{Bu75} to the meson case. In this paper we
just quote the results, both for spin-1 and spin-2 particles. 
Details will be presented elsewhere\cite{RS98}.

{\bf (a) Omega Emission}: 
The omega meson has purely vector interaction, so the matrix 
element for proton to proton-omega emission is:
\begin{equation}
iM_{\omega pp}=<p'\mid -ig_{\omega NN}\ \gamma _{\mu }\mid p>F_{\omega
NN}(K^{2})\epsilon ^{\mu }.
\end{equation} 
After squaring, we get:
$$|M_{\omega pp}|^{2}={1\over 2}g_{\omega NN}^{2}| 
F_{\omega NN}(K^{2})|^{2}Tr\bigl\{ (\pslash'+M_{p})\gamma 
^{\mu }(\pslash+M_{p})\gamma ^{\nu }\bigr\} \big[-g_{\mu \nu 
}+{K^{2}(K^{2}+{m}_{\omega}^{2})\over 
4x^{2}m_{\omega}^{2}P^{4}}k^{\mu }k^{\nu }\big],$$ 
where the last
factor comes from the completeness relation of the omega meson's 
polarization vector, after making the equivalent meson 
approximation. The light-like vector $k$ is given by 
$k=(P,0,0,-P)$, where $P$ is just a frame choice parameter, which
will not appear in the final answer. Taking the small $x$ and 
small $K^2$ limit, we get: \begin{equation} 
\mid M_{\omega pp}\mid ^{2}= 
4 g_{\omega NN}^{2}{K^2 \over x^2}{(K^{2}+m_{\omega }^{2})\over 
m_{\omega }^{2}}|F_{\omega NN}(K^{2})|^{2}.
\end{equation} 
Introducing this result in our expression for the meson 
distribution we finally get:
\begin{equation} 
f_{p\to\omega p}(x,\tilde K^{2})={g_{\omega NN}^{2}\over 4\pi 
^{2}x}\ \intop_{0}^{\tilde K^{2}}{K^{2}\mid 
F(K^{2})\mid ^{2} \over m_{\omega }^{2}(K^{2}+m_{\omega 
}^{2})}dK^{2}. \end{equation} 

{\bf (b) Rho emission}:
The rho meson has a vector interaction term and a 
``helicity-flipping" term. The matrix element for proton to 
proton-rho is:
\begin{equation}
iM_{\rho pp}=\bigl\{ -ig_{\rho NN}<p'\big|\gamma ^{\mu 
}\big|p>-{f_{\rho NN}\over 2M_{p}}<p'\big|\sigma ^{\mu \nu 
}\big|p>q_{\nu }\bigr\} F_{\rho NN}(K^{2}).
\end{equation}
For the proton to neutron-rho case an isospin factor of 
$\sqrt{2}$ should be inserted in the above expression. After 
squaring we get:
\begin{equation}
\big|M_{\rho pp}\big|^{2}={4K^{2}(K^{2}+m_{\rho }^{2})\over x^{2}m_{\rho
}^{2}}\big(g_{\rho NN}^{2}+{f_{\rho NN}^{2}K^{2}\over 4M_{p
}^{2}}\big)\big|F_{\rho NN}(K^{2})\big|^{2},
\end{equation}
and therefore for the rho distribution the following result is 
obtained:
\begin{eqnarray}
f_{p\to\rho _{0} p}&=&{\alpha _{\rho NN}^{(V)}\over \pi x}\intop 
_{0}^{\tilde K^{2}}{K^{2}\over m_{\rho }^{2}(K^{2}+m_{\rho 
}^{2})}\big|F_{\rho NN}{(K^{2})}^{2}\big|dK^2 \nonumber \\
&+&{\alpha _{\rho 
NN}^{(T)}\over 4\pi x}\intop _{0}^{\tilde K^{2}}{K^{4}\over
m_{\rho }^{2}M_{p}^{2}(K^{2}+m_{\rho }^{2})}\big|F_{\rho 
NN}(K^{2})\big|^{2}dK^2, 
\end{eqnarray}
where $\alpha _{\rho NN}^{(V)}=g_{\rho NN}^{2}/4\pi $ and 
$\alpha _{\rho NN}^{(T)}=f_{\rho NN}^{2}/4\pi $. Numerically it 
turns out that the helicity-flipping contribution is much smaller
than the vector part contribution.

The proton can also emit a $\rho^-$ and a $\Delta^{++}$. The 
$p\to \rho^0 \Delta^+$ transition probability is related to the 
previous one by an isospin factor of $2/3$, and the $p\to \rho^+ 
\Delta^0$ by a factor of $1/3$. There are no $\sigma$ or $\omega$
meson coupling to the $p\to \Delta$ transition, because the spin 
of the proton is $1/2$, of the $\Delta$ is $3/2$, while the 
$\sigma$ and $\omega$ have spin zero.

The $N\Delta \rho$ emission matrix element can be written as:
\begin{equation} 
i{f_{\rho N\Delta }\over m_{\rho }}\left[iq_{\mu }<\Delta ^{\nu 
}\mid \gamma ^{5}\gamma ^{\mu }\mid p>\epsilon _{\nu }-iq_{\mu 
}<\Delta ^{\mu }\mid \gamma ^{5}\gamma ^{\nu }\mid p>\epsilon 
_{\nu }\right]t_{\rho N\Delta },
\end{equation} 
where $t_{\rho N\Delta }= 1, {2\over 3}, {1\over 3}$ is the 
isospin factor mentioned above. Using the completeness relation 
for the isobar particle \cite{DKP83}: \begin{equation} 
\sum _{\lambda }u_{\mu }(\lambda ){\bar u}_{\nu 
}(\lambda )={1\over 3}(\pslash'+m_{\Delta })\left[{2\over 
m_{\Delta }^2}p'_{\mu }p'_{\nu }-g_{\mu \nu }-\gamma _{\nu 
}\gamma _{\mu }+{\gamma _{\mu }p'_{\nu }-\gamma _{\nu }p'_{\mu 
}\over m_{\Delta }}\right],
\end{equation} 
and the completeness relation for the $\rho$ in the equivalent 
meson approximation, we get an expression for the emission 
matrix element squared whose value in the small $x$ limit is:
\begin{equation} 
\mid M_{\rho p\Delta }\mid ^{2}={2t_{\rho N\Delta }^{2}
f_{\rho N\Delta }^{2}\over 3m_{\Delta }^{2}m_{\rho
}^{4}x^2}K^{2}(K^{2}+m_{\rho }^{2})\big[K^{4}+(3m_{\Delta
}^{2}+M_{p}^{2})K^{2}\big]\big|F_{\rho N\Delta }\big|^{2}. 
\end{equation} 

So finally the distribution function becomes:
\begin{equation}
f_{p\rightarrow \rho \Delta }(x,\tilde K^{2})={t_{\rho N\Delta 
}^{2} \alpha _{\rho N\Delta }\over 6\pi m_{\Delta }^{2}m_{\rho
}^{4}x}\intop \limits _{0}^{\tilde K^{2}}{K^{4}(K^{2}+3m_{\Delta
}^{2}+M_{p}^{2})\over (K^{2}+m_{\rho }^{2})}\big|F_{\rho N\Delta
}(K^{2})\big|^{2}dK^2
\end{equation} 
where $\alpha _{\rho N\Delta}=f_{\rho N\Delta }^{2}/4\pi $ 
and the isospin factor $t_{N\Delta \rho}^2 = 1,2/3,1/3$ for 
$p\rightarrow \rho ^{-}\Delta ^{++}$, $p\rightarrow \rho 
^{0}\Delta ^{+}$, $p\rightarrow \rho ^{+}\Delta ^{0}$, 
respectively.

{\bf (c) $K^*$ Emission}:
Here the final state can be either a $\Sigma $ or a $\Lambda$. 
Only the latter gives a contribution comparable with those of the
$\omega$ and $\rho$ cases. In the same way as before we get:
\begin{eqnarray}
f_{p\to K^*\Lambda}&=&{\alpha _{K^*N\Lambda}^{(V)}\over \pi 
x}\intop _{0}^{\tilde K^{2}}{K^{2}\over m_{K^* 
}^{2}(K^{2}+m_{K^*}^{2})}|F_{K^*N\Lambda}{(K^{2})}|^{2}dK^2 
\nonumber \\ &+&{\alpha
_{K^*N\Lambda}^{(t)}\over 4\pi x}\intop _{0}^{\tilde 
K^{2}}{K^{4}\over m_{K^*}^{2}M_{p}^{2}(K^{2}+m_{K^* 
}^{2})}\big|F_{K^*N\Lambda}(K^{2})\big|^{2}dK^2, 
\end{eqnarray}
where $\alpha _{K^*N\Lambda}^{(V)}=g_{K^*N\Lambda}^{2}/4\pi $ and
$\alpha _{K^*N\Lambda}^{(T)}=f_{K^*N\Lambda}^{2}/4\pi $. 

\subsection{Spin 2 meson distributions}

These contributions are progressively smaller due to the higher 
mass of the spin 2 particles. Thus we will only consider the 
lightest spin 2 particle, the $f_2(1270)$, and we will leave its 
coupling constant as an adjustable parameter.

The $f_2 NN$ matrix element can be written as:
\begin{equation} 
iM_{f_2NN}\ =\ {if_{f_2NN}\over 4m_{f_2}}<p'\mid V^{\mu \nu 
}\mid p>\phi _{\mu \nu }F_{f_2NN}(K^{2}),
\end{equation} 
where the symmetric tensor $\phi _{\mu \nu }$ is the spin-2 
particle field. The vertex $V^{\mu\nu}$ is\cite{vertex}:
\begin{equation} 
V^{\mu \nu }= {4M_p \over s^2} s^{\mu }s^{\nu }+(s^{\mu
}\gamma ^{\nu }+s^{\nu }\gamma ^{\nu }),
\end{equation} 
with $s^{\mu }=p^{\mu }+p'^{\mu }$.

After squaring, averaging over initial spins and summing over 
delta spins, we get: 
$$\mid M_{f_2NN}\mid 
^{2}={f^{2}_{f_2NN}\over 32 m^2_{f_2}}\Lambda _{\mu \nu ,\rho 
\sigma }Tr\{(\pslash'+m)V^{\mu \nu }(\pslash+m)V^{\rho\sigma}\} \
F_{f_{2}NN}(K^{2}),$$ 
where the $\Lambda _{\mu \nu ,\rho \sigma 
}$ factor comes from the equivalent particle approximation, and 
is given by: 
\begin{eqnarray} 
\Lambda _{\mu \nu ,\rho \sigma }&=&{1\over 2}(g_{\mu \rho }g_{\nu
\sigma }+g_{\mu \sigma }g_{\nu \rho })+{1\over 2}{K^{2}\over 
(2xP^{2})^{2}} \nonumber \\
&\times&( g_{\mu \rho }k_{\nu }k_{\sigma }+g_{\mu
\sigma }k_{\nu }k_{\rho }+g_{\nu \rho }k_{\mu }k_{\sigma }+g_{\nu
\sigma }k_{\mu }k_{\rho })  \\ 
& +&{K^{4}\over 
(2xP^{2})^4}\left({1\over 2}-{2K^{2}\over m_{f}^{2}}+{K^{4}\over 
m_{f}^{4}}\right)k_{\mu }k_{\nu }k_{\rho }k_{\sigma } \nonumber. 
\end{eqnarray} 
Thus, in the small $K^2$ and small $x$ limit we get:
\begin{equation} 
|M_{f_2NN}|^2={4f_{f_{2}NN}^{2}\over m_{f_{2}}^{2}}{K^{4}\over 
x^{4}}\big({1\over 2}-{2K^{2}\over m_{f}^{2}}+{K^{4}\over
m_{f}^{2}}\big){(16M_{p}^{2}+K^{2})\over
(4M_{p}^{2}+K^{2})}|F_{f_{2}NN}(K^{2})|^{2}, 
\end{equation} 
and therefore we obtain for the $f_2$ distribution function:
\begin{equation} 
f_{p\rightarrow f_{2}p}={f_{f_{2}NN}^{2}\over 4\pi 
^{2}m_{f}^{2}x^{3}} \intop \limits _{0}^{\tilde
K^{2}}{K^{4}(16M_{p}^{2}+K^{2})\over
(4M_{p}^{2}+K^{2})(K^{2}+m_{f}^{2})^{2}}\big({1\over 2}-{2K^{2}\over
m_{f}^{2}}+{K^{4}\over m_{f}^{4}})|F_{f_{2}NN}(K^{2})|^{2} dK^{2}. 
\end{equation}

\subsection{Parton Distributions in Mesons} 

 From the above calculations we have obtained various distribution
functions $f_{m/p}(x,Q^2)$ in the equivalent meson approximation.
In order to compare our results in Eq. (1) with the experimental 
data from HERA\cite{Zeus95}, we need the parton distribution 
functions $F_2^m(\beta,Q^2)$ for each of the mesons we have 
considered ($\pi^0,\pi^+,\rho^-,\rho^0,\rho^+,K^{*+},f_2)$.

The parton content of mesons is presently poorly known. Thus we 
will assume here that the valence, sea and strange quark 
distributions of all the mesons mimic the parton distributions of
pions. For the pion's distribution functions, we use the GRV 
parameterization\cite{GRV92}. Then each of the $F_2^m(x,Q^2)$ can
be written in terms of three independent functions: (1) the 
valence quark distribution: $x v^{\pi}(x,Q^2)$, (2) the light 
sea-quark distribution: $x\bar{q}^{\pi}(x,Q^2)$, and (3) the 
strange sea-quark distribution: $x\bar{s}^{\pi}(x,Q^2)$. That is:
\begin{equation}
F_{m}^{2}(x,Q^{2})=C_{1}^{m}xv^{\pi }(x,Q^{2})+C_{2}^{m}x\bar
{q}(x,Q^{2})+C_{3}^{m}x\bar {s}(x,Q^{2}).
\end{equation}
The coefficients for the charged mesons $m=(\pi^+,\rho^-,\rho^+)$
are $(C_1^m,C_2^m,C_3^m)=({5\over 9},{10\over 9},{2\over 9})$, 
and for the neutral mesons $m=(\pi^0,\rho^0,f_2)$ are $(C_1^m,C_2^m,C_3^m)= 
({5\over 18},{10\over 9},{2\over 9})$. For the special case of 
the strange meson $K^{*+}$, for simplicity we will assume a 
parton distribution $F_{m}^{2}(x,Q^{2})$ similar to that of the 
charged mesons.

\section{\bf Results} 

There are three parameters in our model, namely, the momentum 
squared cut-off $\tilde{Q}^2$ (which should be around 
$\Lambda^2_{QCD}$), the tensor-meson coupling constant 
$\alpha_{f_2NN}={f_{f_2NN}^2\over 4\pi}$, and the form factor 
momentum cut-off scale $\Lambda_{f_2NN}$. We have chosen the 
values $\tilde{K}^2=0.048 [GeV^2], f_{f_2NN}=4$, and 
$\Lambda_{f_2NN}=1.29 [GeV]$.

\vspace{0.5cm}
\begin{figure}[htb]
\begin{center}
\leavevmode {\epsfysize=13cm \epsffile{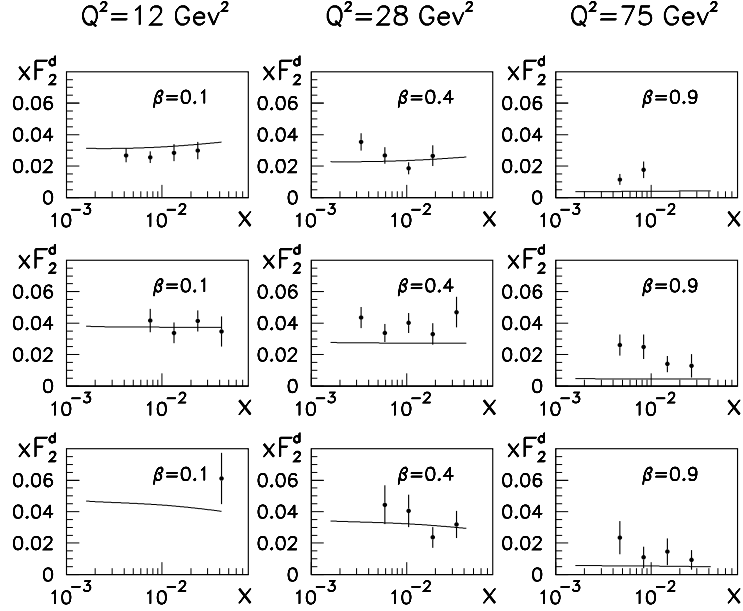}} 
\end{center}
\caption[*]{\baselineskip 13pt 
Comparison between the predictions of the peripheral meson model 
and experimental data\cite{Zeus95}.} \label{usmth76f2} 
\end{figure}  
 
In Fig. 2 we compare the results of our model with the 
experimental data from HERA\cite{Zeus95}. We have plotted the results 
for $F_2^d(x,Q^2,\beta)$ for $Q^2=12,28,75 [GeV^2]$ and $\beta 
=0.1,0.4,0.9$. Form the plots we can see that the x-dependence of
$F_2^d(x,Q^2,\beta)$ can be explained from the x-dependence of 
the equivalent meson content of the proton.

\section{\bf Conclusions} 

We have presented here an equivalent-meson model in order to 
explain the observed large rapidity gap events at HERA. We have 
seen that the peripheral meson content of the proton can explain 
the x-dependence of the measured $F_2^m(x,Q^2,\beta)$ structure 
functions. Thus, from standard nuclear physics knowledge and the 
parton distribution of mesons, the essential features of this 
class of events can be explained. Our model implies the 
existence of interesting final states in the forward baryon, 
including helicity-flipping proton and isospin-changing states 
(e.g. $\Delta^{++}$) in the forward direction, which should be 
interesting to observe in large rapidity gap events.

 \bigskip
{\bf Acknowledgments:} We would like to thank Stanley J. Brodsky 
for helpful conversations.  This work has been 
partially supported by Fondecyt (Chile) under grant
1990806 and by a C\'atedra Presidencial (Chile). R. Rivera thanks
Fundaci\'on Andes for a Doctoral grant.


\begin{thebibliography}{99}

\bibitem{Meson}

See for example: R.~Machleidt, in ``Advances in Nuclear 
Physics'', Vol. 19, 189 (Plenum Press, 1989).

\bibitem{Zeus95}

M.~Derrick et al., Zeits.~f\"ur~Phys.~{\bf C68}, 569 (1995);
C.~Adloff et al., Zeits.~f\"ur~Phys.~{\bf C76}, 613 (1997).

\bibitem{W97}

See for example: M.~W\"usthoff, Phys.~Rev.~{D56}, 4311 (1997),  
and references therein.

\bibitem{GRV92}

M.~Gl\"uck, E.~Reya and A.~Vogt, Zeits.~f\"ur~Phys.~{\bf C53}, 
651 (1992); M.~Gl\"uck, E.~Reya and I. Schienbein, 
preprint DO-TH 99/01, hep-ph/9903288.

\bibitem{BH89}

B. Holzenkamp et al., Nucl.~Phys.~{\bf A500}, 485 (1989).

\bibitem{Bu75}

See for example: V.~M.~Budnev et al., Phys.~Rep.~{\bf C15}, 181 
(1975).

\bibitem{RS98}

H.-J. Lu, R.~Rivera and I.~Schmidt, in preparation.

\bibitem{DKP83}

O.~Dumbrajs et al., Nucl.~Phys.~{\bf B216}, 277 (1983).

\bibitem{vertex}

K.~V.~Vasavada, Phys.~Rev.~{\bf D9}, 1918 (1975).

\nonfrenchspacing
\end{thebibliography}
\end{document}